\documentclass[a4paper,conference]{IEEEtran}
 \usepackage{graphicx} 
 \usepackage[cmex10]{amsmath} 
 \usepackage{amssymb,verbatim}

\begin{document}
%
\title{%
Comparison of security mechanisms of Mathematical cipher, Wyner scheme, QKD, and Quantum stream cipher\\
}

\author{
\IEEEauthorblockN{Gikyu Yamamoto$^{1}$, Osamu Hirota$^{1,2,3}$ \\}
\IEEEauthorblockA{
1.Optical Quantum Information Technology Laboratory\\
Totsuka-ward, Yokohama, Kanagawa Pref. Japan\\
2. Quantum ICT Research Institute, Tamagawa University\\
6-1-1, Tamagawa-gakuen, Machida, Tokyo 194-8610, Japan\\
3. Research and Development Initiative, Chuo University, \\
1-13-27, Kasuga, Bunkyou-ku, Tokyo 112-8551, Japan\\
{\footnotesize\tt E-mail:yyjj0655@gmail.com,  hirota@lab.tamagawa.ac.jp, hirotaosamu01@gmail.com} \vspace*{-2.64ex}}
}

\maketitle

\begin{abstract}
A new generation of global communications technology has been emerging. These systems, which utilize established device technologies 
and quantum effect devices, require ultra-high speeds,  low cost, and strong security. 
In recent years, global communication systems have faced various practical security challenges depending on their configurations, 
and research efforts are underway to address these issues.
In particular, the issue of the security of  physical layer security from microwave wireless systems to quantum optical communication systems 
is urgent problem.
However, concepts of cryptographic schemes have also been diversifying. Typical examples are mathematical ciphers, the Wyner scheme and QKD. 
Then, the Y-00 protocol has recently emerged as a third pillar cryptographic technology in the optical quantum domain. 
These security principles differ significantly from one another.
This makes it difficult for different fields to understand each other.
At this stage, comparative explanations of the security principles underlying these various cryptographic technologies are likely to 
promote mutual understanding among researchers across different fields. As the first trial, 
this lecture note explains the security mechanism of the third pillar (Y-00), comparing it with the principles of other mechanisms.

\end{abstract}

%
\IEEEpeerreviewmaketitle

\section{\textbf{Introduction}}

Since Shannon suggested the direction of cryptography in its early days, modern mathematical cryptography has made remarkable progress. 
Shannon attempted to establish a new cryptography system based on his own theory of information and communication, concluding that 
the only information theoretic secure cryptography considered effective for data encryption was One time pad (OTP). 

However, OTP's mechanism is far removed from the convenience of cryptography, prompting mathematical scientists to shift 
to mathematical cryptography [1]. This system uses a short secret key and mathematically extended pseudorandom numbers. 
Security then became a focus of computational complexity based security. It is clear that exponential computational complexity poses 
no practical problems. 

Meanwhile, with the globalization of communications, the issue of key distribution for data encryption scheme became increasingly important, 
leading to the invention of $\textbf {Public key cryptography (PKC)}$, which has become an essential technology for 
today's Internet infrastructure [2]. 
This was also due to the establishment of the framework of computational complexity security.

\begin{figure}
\centering{\includegraphics[width=8cm]{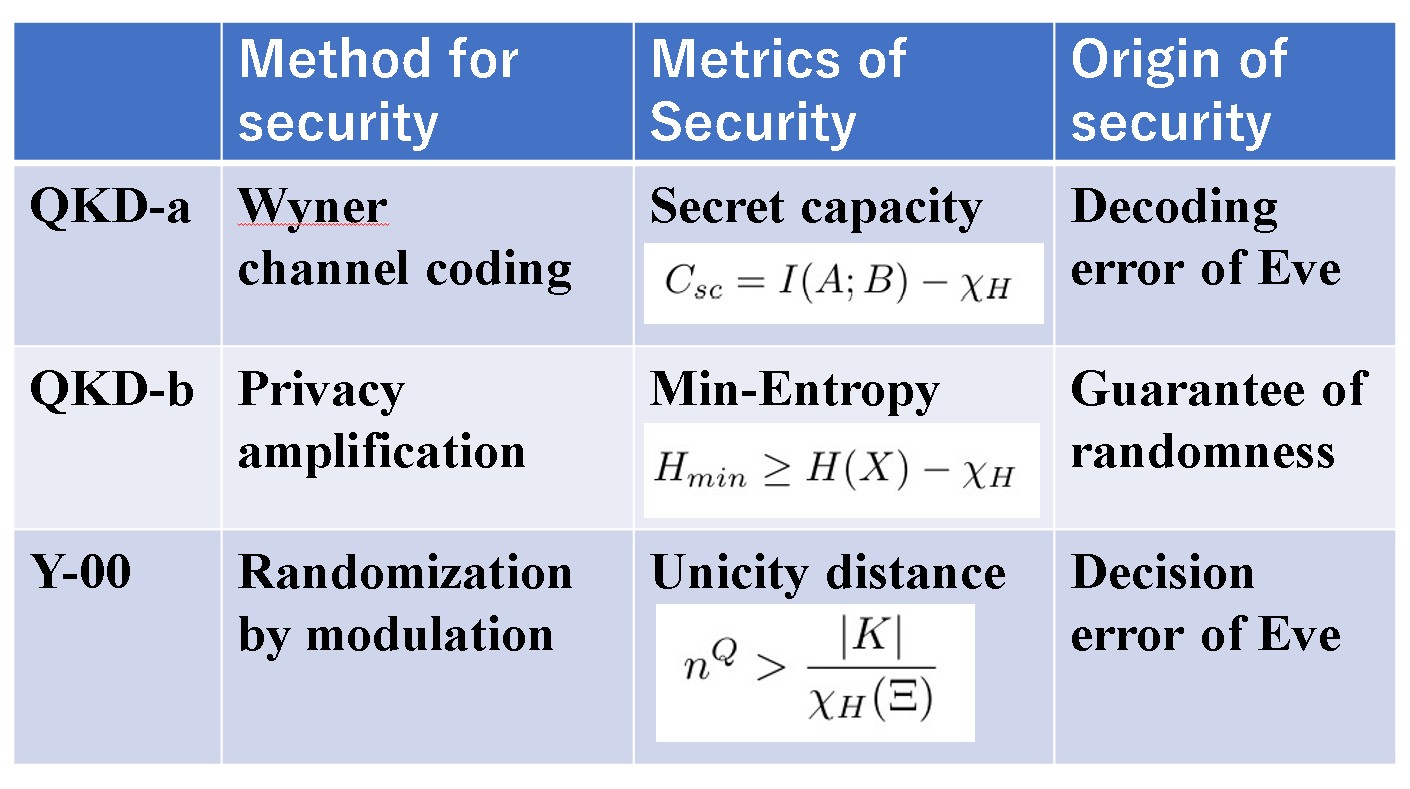}}
\caption{\textbf {Summary of comparison of security mechanisms of early and current QKD, and quantum stream cipher (Y-00). }
$\chi_H$ is the Holevo capacity, $\chi_H(\Xi)$ is the Holevo capacity as function of randomization by modulation ;$\Xi$. 
The purpose of this note is to provide an explanation of this diagram. A detailed comparison is discussed in the appendix.}
\end{figure}

In the 21st century, quantum computing technology has presented challenges for cryptography. Nevertheless, mathematical cryptography has 
shifted to research on quantum-safe cryptography, which has reached a practical level of security while still ensuring convenience as 
a fundamental technology for the scale of the Internet. On the other hand, information-theoretically secure cryptography for data encryption 
and key distribution is attractive, and so research on this field, although minor, seems to have continued.

For a long time, research into information-theoretically secure cryptography was conducted under the concept of 
$\textbf {Wyner's wire-tap channel scheme}$, 
which exploits the noise characteristics of the communication channel. This has the advantage of eliminating the need for a shared secret key.
However, it requires the assumption that the eavesdropper's communication channel is noisier than the legitimate receiver's one. 
It is called Wyner condition.
The parameter that guarantees security is the secret capacity, which is defined as the difference between the capacity of the legitimate 
 channel and the capacity of the channel to an eavesdropper.
If this parameter is positive, then according to the Shannon coding theorem, a system can be realized in which an eavesdropper cannot 
correctly receive encoded information.

Recently, there has been much research into the practical implementation of such models in microwave wireless communication systems [3], 
which is called  $\textbf {Physical layer security (PLS)}$. 
Its main technical challenges are the technology to create the Wyner condition and the theory of codes for finite systems, 
which departs from asymptotic theory. 
 Furthermore, this technique may provide information-theoretically secure key distribution.
Research for key deistribution was also conducted to create the Wyner condition of wire-tap channel. 
However, it requires stringent performance requirements, such that it is necessary to ensure the randomness of the shared sequence.
At that time, the evaluation method was converted from mutual information to statistical distance, 
and randomness assurance techniques were developed (See appendix).

On the other hand, the development of  the scheme which utilizes quantum mechanical effects has been proposed.
It is  called  a $\textbf {Quantum key distribution(QKD)}$. 
The development for the first invention as BB-84 began with a quantum version of the Wyner model [4]. 
However, it did not guarantee sufficient security, and so the conversion from quantum coding to a privacy amplification mechanism based on Renner theory [5], 
and improvement methods such as decoy method and device-independent methods have led to  information-theoretically secure scheme.

Regarding the symmetric key cipher with the information theoretic security for high-speed data encryption method, the OTP is so inconvenient. 
Thus the issue of information theoretic security considered that little progress has been made. 

In 2000, a third pillar of technology has emerged in cryptography [6].  This involves ensuring the security of cryptographic 
functions by mechanism of communication system. Specifically, the concept is to randomize the modulation mechanism in optical 
communication systems, thereby increasing the quantum noise effect on the randomized modulation signal (ciphertext) that an eavesdropper 
can receive compared to that of a legitimate receiver with a key. This encryption method is called $\textbf {Quantum stream cipher (Y-00)}$. 
This research began in 2002 [7] and 2003 [8] with the public release of experimental equipment development.

The security of the entire system depends on the extent to which the data (plaintext) and secret key information  (running key sequence) are 
masked by quantum noise. 
Thus, the security of a cryptosystem is determined not only by the randomness of the key but also by the nature and amount of quantum noise, 
making it outside the scope of the Shannon impossibility theorem [9]. It means  that data of any length can be encrypted in an information-theoretically 
secure manner even if a cipher based on a short secret key is used. 
Breaking the Shannon limit in this way asserts that there exists a cipher that does not satisfy the prerequisites of its theorem. 
In mathematical science, this situation is called a lifting.

However, research into the Y-00 type is still ongoing. One of reasons is that the quantum noise level in the first-generation 
standard quantum stream cipher is small. Therefore, while data is completely masked, the running key information is not sufficiently masked. 
This means that data security is unconditionally guaranteed, but key security is not perfect. 

Meanwhile, in recent years, a method for extending the quantum noise effect proposed over a decade ago [10] has been reevaluated, 
and it has been pointed out that  it achieves sufficient information-theoretic security against the attack of secret key [11].  
This method does not impose any conditions on the cryptographic properties of the mathematical encryption component.  
This kind of scheme is called a generalized quantum stream cipher.

In light of the above, this lecture note aims to explain the most fundamental theoretical background of the three cryptographic technologies 
of information theoretic security as concisely as possible, with comparisons and focus on the emergence of the third pillar (Y-00). 
We hope that this paper will clarify the meaning of what is shown in Fig.1.

\section{\textbf {Theory  for conventional cryptography }}

Let us give a backgound on the theory of the conventional cipher in this section [1,2].

\subsection{\textbf {Shannon's scheme of cryptography}}
It is well known that in 1949, Shannon published a paper at the BSTJ in which he attempted to clarify the basic concepts of cryptography 
by applying his own entropy-based information theory. First, let's establish the cryptographic mechanism under consideration. 
Let $X$ be the plaintext sequence 
and $ K$ be the secret key. The ciphertext is given by
\begin{equation}
Y =  F(X, K) 
\end{equation}
where $F$ is the encryption function. The secret key includes either a running-key sequence obtained from a PRNG with a short initial key, 
or a truly long random number sequence.
First, let's state the basic assumptions of Shannon's theory. Shannon's theory assumes that when constructing an encryption mechanism, 
the ciphertext $Y^A$ is set by a legitimate sender and can be correctly received by both the legitimate receiver and an eavesdropper. 
That is,
\begin{equation}
Y^A = Y^B = Y^E \equiv Y
\end{equation}
Here, $Y^B$ and $Y^E$ are ciphertext sequences that can be received by both the legitimate receiver and an eavesdropper. 
This mechanism can be expressed as follows using Shannon's entropy theory:
\begin{equation}
H(X|K,Y^B) = H(X|K,Y^E) = 0 
\end{equation}
That is, the plaintext can be completely decrypted with the ciphertext and the key.\\

\subsection{\textbf {Shannon impossibility theorem and One-Time Pad encryption,}}

On the other hand, for cryptosystems that satisfy the condition in Eq(2), the following theorem holds:\\

$\textbf {Theorem 1}$: \\
The security of plaintext (data) based on ciphertext only attack of encryption scheme is limited as follows:
\begin{equation}
H(X|Y) \le H(K)
\end{equation} 
where $X$ is plaintext, $Y $ is ciphertext, and $K$ is secret key, respectively.\\

This theorem is called the Shannon limit or Shannon impossibility theorem. Below, we explain the specific characteristics of Shannon cryptosystems 
that satisfy the condition in Eq(2).
Prepare a key sequence K for encrypting plaintext sequence X. Assume that the ciphertext consists of these two sequences. 
In this case, a conventional cryptosystem has the following property:\\

$\textbf {Definition 1}$: Cryptosystem whose constructed ciphertext satisfies the following property is called a non-random cryptosystem.
\begin{equation}
H(Y^E|X,K)=H(Y^B|X,K)=0
\end{equation}

A typical example of this type of cryptosystem is an additive stream cipher. In this case, 
Shannon defines perfect secrecy in the above cryptographic mechanism as follows:\\

$\textbf {Definition 2}$: A cryptosystem has perfect secrecy if it satisfies the following property:
\begin{equation}
H(X|Y)=H(X)
\end{equation}

An encryption mechanism that satisfies the above conditions is called a one-time pad (or Vernam cipher). It is inefficient and places 
an excessive burden on key management. In other words, sharing an inexhaustible key to guarantee information-theoretically secure encryption 
is technically difficult to achieve. 

\subsection {\textbf {Randomized encryption}}

A cipher which involves randomization in the encryption process is called a randomized encryption [27]. 
 In this subsection, we shall explain the concept. There are two types of schemes as follows.

\subsubsection{\textbf {Wyner scheme for information theoretics security}}

By focusing on coding within the same Shannon information theory, Wyner proved that information-theoretically secure encryption can be 
achieved without a shared key. 
Because this does not use a key, it falls outside the scope of the Shannon impossibility theorem. 
However, for this to be true, the SNR of the signal received by the eavesdropper must be lower than that of the legitimate receiver. 
This is called the Wyner condition.
As a result, the secret capacity as a security metrics is defined as the difference between the mutual information of Alice-Bob channel and 
the capacity of Alice-Eve channel. 
Then, the  decoding error of Eve based on Shannon channel coding theorem assures the security, when the legitimate communicators adopt 
Wyner wire-tap coding, or coset coding.
This method will lead to developments in physical layer security in wireless systems and QKD.

\subsubsection{\textbf {Random cipher with private randomization}}

In general, symmetric key cipher for data encryption is required to be more robust. Research to improve the security of symmetric key cipher continues. 
An example is  a cipher, which inherently incorporates the concept of private randomization. It is called a random cipher.
In Shannon theory, it is possible to conceive of cryptography that is information-theoretically secure for a secret key by means of random cipher, 
even if it does not satisfy the complete confidentiality condition. Here, we provides a definition of random cipher.\\

$\textbf {Definition 3}$. A cryptosystem is called a random cipher if it has the following property:
\begin{equation}
H(Y|X,K)\ne 0
\end{equation}

Specifically, its information-theoretic security  on key is evaluated using the unicity distance. 
However,  a security on data of a cipher with this structure is subject to the constraints of the Shannon impossibility theorem, 
since its encryption mechanism satisfies the condition of Eq(2).\\

\section{\textbf{Concept of generalized random cipher}}

In this section, we explain the main concept of  the  generalized random cipher.

\subsection{\textbf {Conceptual mechanism}}
In general, the discussion on the information theoretic security of the symmetric key cipher is discussed based on one time pad (OTP)
under  the Shannon impossiblity theorem.
OTP provides perfect security against ciphertext-only attacks on the plaintext.  
Since the key sequence has the added of being permanently replenished, 
which also guarantees security against known-plaintext attacks on the key.
However, the one-time pad is not a random cipher, so it is not perfect against tampering attacks or other potential attack on data such as 
a partial known plaintext attack following a statistical information for the plaintext.
 
In 2000, the concept of generalized random cipher was proposed for 
eliminating the drawbacks of OTP mentioned above [6]. 

Here let us define a generalized random cipher as a cryptosystem such that the ciphertext of the cryptosystem is hidden by ideal noise, 
and it has the following features.
\begin{equation}
Y^A=Y^B \ne Y^{Eq}
\end{equation}
where $Y^A$ and $Y^B$ are the ciphertexts of the legitimate communicator, $Y^{E_q}$ is the ciphertext with noise received by the eavesdropper.
In other words, this is achieved by creating a situation where the ciphertext received by the legitimate receiver and the ciphertext that 
can be received by the eavesdropper are different.
If we describe this as a sequence, it is as follows:
\begin{eqnarray}
&&Y^B_n=y^A_1, y^A_2, y^A_3, \dots, y^A_n \\
&&Y^{Eq}_n =y^{E_q}_1,y^{E_q}_2,y^{E_q}_3, \dots, y^{E_q}_n\nonumber \\
&&=y^A_1\oplus q_1, y^A_2\oplus q_2, y^A_3\oplus q_3, \dots, y^A_n\oplus q_n 
\end{eqnarray}
where $q_1,q_2,q_3, \dots$ denote errors due to true noise.
If this situation could be realized, then the following conceptual relations would be held.
\begin{eqnarray}
&&H(Y^B_n|K,X_n)=0 \\
&&H(Y^{E_q}_n|K,X_n) \ne 0
\end{eqnarray}
where $Y^B_n,Y_n^{E_q}$ are ciphertext for Bob and Eve, respectively, $K$ means secret key
 and $X_n$ is plaintext.

 That is, the ciphertext is randomized by noise.
The important features of the generalized random cipher are as follows:
Even with a cryptographic mechanism that uses a short secret key and PRNG, it is still possible to provide information theoretic security 
against  ciphertext only attacks and known plaintext attack. 
These properties mean that the Shannon impossibility theorem for data security in the cryptology can be lifted.

\subsection{\textbf{Security  against ciphertext only attack on data}}

Due to the Shannon impossibility theorem, the data security against ciphertext only attack is in general limited as shown in the theorem 1 when it holds
\begin{equation}
Y^A=Y^B = Y^{Eq}
\end{equation}
However, a generalized random cipher does not meet this requirement from Eq(8).
Therefore, it does not need to obey the Shannon impossibility theorem in a ciphertext-only attack. 
Thus, it is said that the Shannon impossibility theorem is lifted, if the cipher has the following property:
\begin{equation}
H(X|Y^{E_q}) > H(K)
\end{equation}

\subsection{\textbf{Generalized unicity distance theory for key security}}
In general, the performance of the information theoretic security of the symmetric key ciphers on key can be evaluated 
by the unicity distance or spurious key. 
Thus, we need the unicity distance theory  to evaluate key security of the generalized random cipher.. 
It is called a generalized unicity distance.
The detailed discussions on the generalized unicity distances for such a cryptographic mechanism have been given in [9, 12]. 
The following is the short summary for those discussion.\\

\subsubsection{\textbf{Ciphertext only attack}}

The unicity distance of a ciphertext only attack on key is defined for the eavesdropper's ciphertext as follows:\\

$\textbf {Definition 4}$: \\
Let $n^Q_0$ be the minimum length of the ciphertext that has zero key ambiguity for the eavesdropper's ciphertext.
Then it is given by
\begin{equation}
n^Q_0: H(K|Y^{Eq}_{n^Q_0})=0
\end{equation}
 $n^Q_0$ is called the unicity distance of ciphertext only attack for generalized random cipher.

Unlike the conventional type, the above equation does not necessarily depend on the statistical structure of the plaintext,  
but on the randomness of the ciphertext with noise that can be obtained by the eavesdropper.\\

\subsubsection{\textbf{Known plaintext attack}}

The conventional random ciphers can achieve a large unicity distance for a ciphertext-only attack.
However, it is impossible to guarantee information-theoretic security more than a key length in the known plaintext attack.

Here, we consider a known-plaintext attack on generalized random cipher. The difinition of the unicity distance for known plaintext attack
  is given as follows:\\

$\textbf {Definition 5}$:\\
The unicity distance of known plaintext attacks for generalized random cipher is defined as follows:
\begin{equation}
n^Q_1: H(K|X_{n^Q_1}, Y^{Eq}_{n^Q_1})=0
\end{equation}

In the case of generalized random cipher, at least, the following can be expected.
\begin{equation}
|K| \ll n^Q_1 \le 2^{|K|}
\end{equation}
where $|K|$ is the key length. Such a possibility arises from the nature of Eq(12).
This performance is the most important in the practical applications.
One should emphasize that this property is not possible with existing cryptography theory.
Thus, the next challenge is how to realize such a mechanism.

\section{\textbf{Quantum stream cipher as realization of generalized random cipher}}
Let us explain the sutructure of quantum stream cipher in this section.

\subsection{ \textbf {Conceptual explanation of quantum stream cipher} (Combining Communication methods, Quantum effects, and Cryptography)}

Generally, the security of symmetric key cryptography is determined by the properties of the key sequence. 
Therefore, the scientific challenge is to devise ways to approximate the key sequence to true randomness. 
The Y-00 protocol uses a cryptographically secure pseudorandom key sequence, but this is used to degrade 
the eavesdropper's reception accuracy (e.g., SNR), and the randomness of the key sequence is considered secondary. 

In other words, the primary encryption mechanism is a mechanism that increases the difference in reception accuracy between the legitimate receiver and the eavesdropper. In particular, the quantum communication theory and quantum Shannon information theory perspective, which states that the quantum noise 
characteristics that determine the eavesdropper's reception accuracy can be controlled by signal structure and modulation method, are essential. 
 
 Thus, in Y-00, the communication method itself is the encryption mechanism, 
 and the security of the system is determined by the amount of quantum noise experienced by the eavesdropper. 
 This is a concept that has not been adopted in previous cryptography and as a result the Shannon impossibility theorem is lifted.

\subsection{\textbf {Mechanism for making difference of reception capabilities between Bob's and 
Eve's receivers using optical quantum modulation}}

Here, we explain the structure of an encryption mechanism using modulation technology. 
First, $M$ communication bases for binary phase modulation are prepared. 
One of the $M$ communication bases is randomly selected using mathematical cryptography. 
Using the selected communication base, the binary plaintext is transmitted using binary phase modulation. 
Because the communication base used for each bit of the plaintext sequence is selected using mathematical cryptography,
 the binary plaintext for each signal appears to be randomly spread across $2M$ signals [6,7,8] . 

Specifically,  the signal configulation as quantum ciphertext is that $2M$ signals  are arranged on a circle with amplitude A on the phase plain. 
And, the adjacent 0s and 1s in each signal correspond to the plaintext. 

Because the legitimate receiver can always receive binary phase modulation using the basis information for each slot based on key information, 
errors are almost negligible and the plaintext is received correctly. 
 Since the eavesdropper does not know the key, they have to adopt a reception method for $2M$-value signals  (See Fig.2 and Fig.3). 
Details of this will be explained later.

\begin{figure}
\centering{\includegraphics[width=8cm]{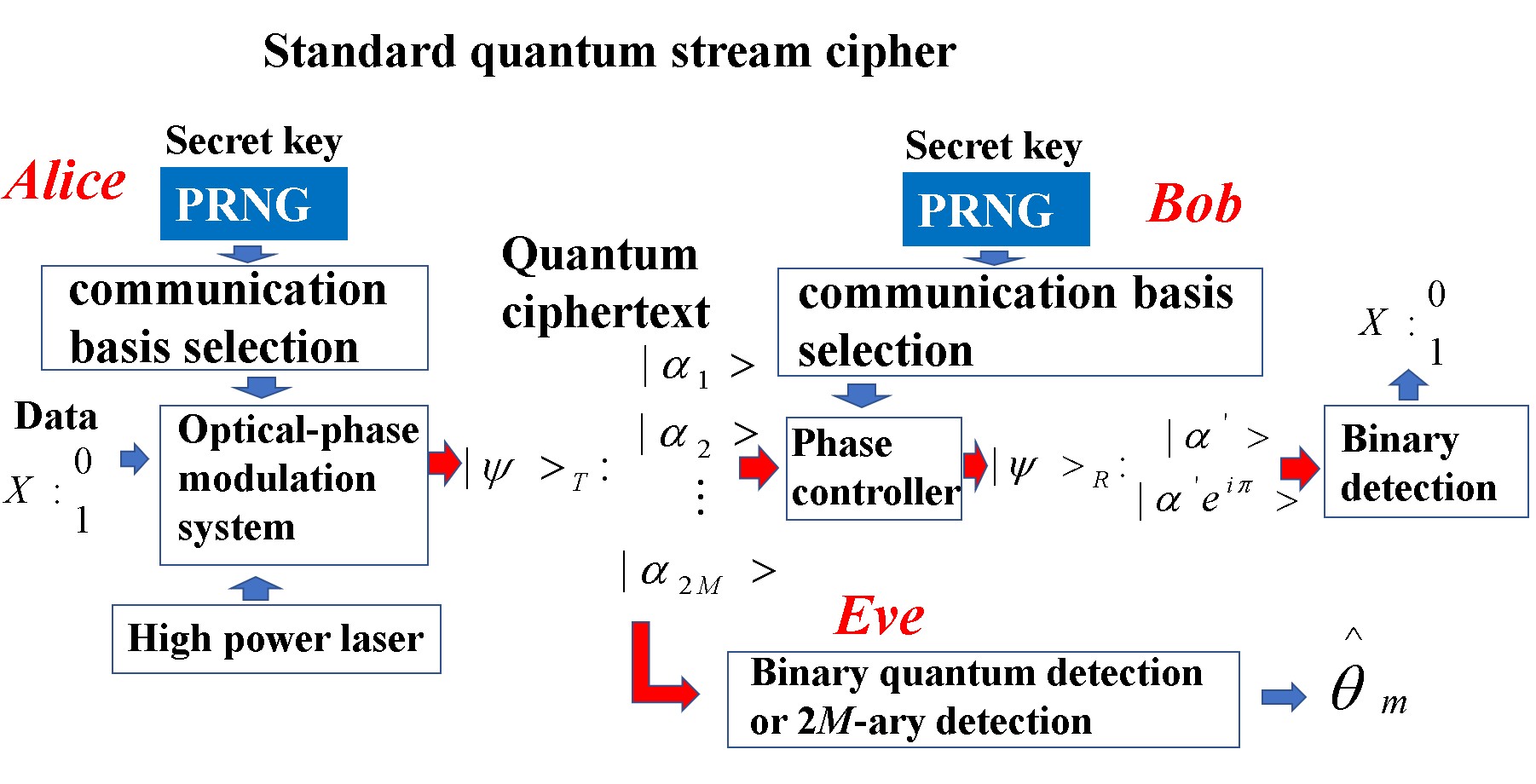}}
\caption{\textbf {Scheme of encryption and decryption of standard quantum stream cipher.}
The optical modulator system selects one  basis from a set of $M$ communication bases according to a running key. 
It is done by phase controller. Then it performs binary PSK for plaintext (data) by using selected basis.}
\end{figure}

\begin{figure}
\centering{\includegraphics[width=5cm]{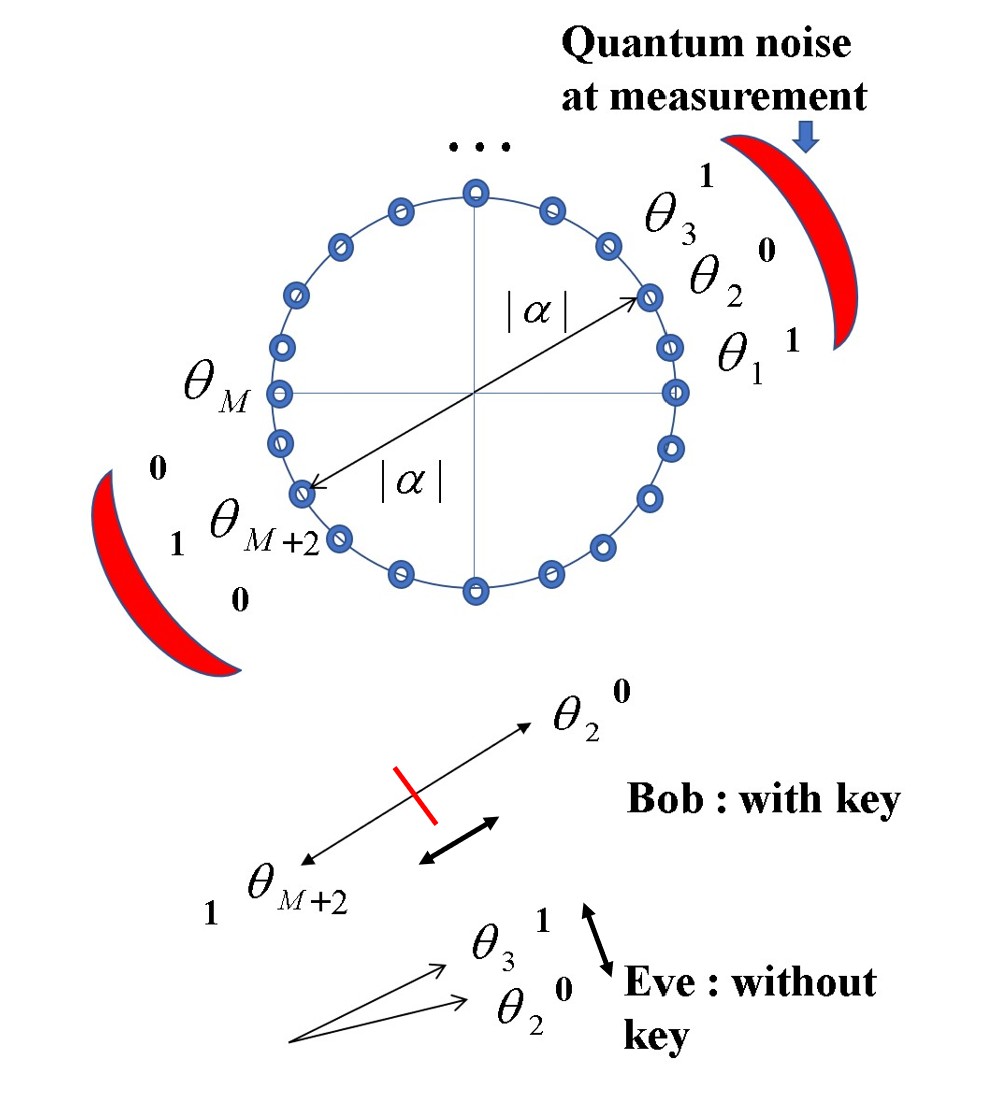}}
\caption{\textbf {Quantum noise effect to Bob and Eve on plaintext. }
Upper part shows the phase signals on phase plain of light wave which correspond to the physical ciphertext.
The phase reference frame of Bob's receiver is adaptively rotated by the running key sequence. 
So Bob's decision becomes always $+\alpha$ or $-\alpha$. Because she does not know the secret key, 
 Eve has to adopt  a receiver for $2M$ value, and she has to identify adjacent signals hidden by quantum noise. 
 In order for Eve to obtain the key information, she needs to identify the exact value of the phase signals 
that correspond to the running key.}
\end{figure}

\subsection{\textbf {Assumptions of eavesdropper's attack capability }}
In any physical scheme of encryption, the ability of quantum measument to obtain information from quantum states which convey 
the data should be defined, and the eavesdropper can adopts any quantum measurement allowed by quantum mechanics.

In the Y-00 protocol, there is only one communication channel from Alice to Bob, so we assume that an eavesdropper can adopt 
any measurement or reception method allowed by quantum mechanics for that channel. In QKD, the existence of a classical communication channel 
for  information exchange makes security extremely complex, and as will be discussed later, comparing the two is prone to misunderstandings.

First, the method in which an eavesdropper applies quantum optimal reception to each signal slot is called a quantum individual measurement. 
Second, the method in which quantum state signals present in multiple slots are entangled and then measured all at once is called 
a quantum collective measurement. 
Thus,  we assume that Eve can adopt following quantum measurements.\\
(a) Quantum Individual Measurement, (b) Quantum Collective Measurement.

Here, we have the following theorem to discuss the difference between QKD and Y-00.\\

\textbf {Theorem 2}: \\
Achieving Holevo capacity requires encoding of quantum state signals and quantum collective measurement of the encoded signals.\\

The difference in the effectiveness of quantum collective measurement in QKD and 
Y-00 protocol is explained by this theorem. 

In QKD, because there is a classical communication channel, it is possible to store the data in a quantum memory and 
then perform an attack in which the error-correcting coded quantum state sequence is quantum-measuring all at once, 
and the Devetak-Winter theorem states that the maximum amount of information at that time is the Holevo capacity.

On the other hand, in Y-00 protocol, there is no classical channel and the quantum state signal is not coded, 
resulting in uncoded quantum collective measurement. 
The quantum gain in this case is small and its role is insignificant. 
However, such theory can be used to estimate  an upper bound on the mutual information. 
Furthermore, it is noteworthy that Y-00  protocol has the property that the mutual information (or Holevo capacity) of an eavesdropper can be reduced 
by the modulation method with additional randomizations. 

In QKD-a, the mutual information and Holevo capacity of an eavesdropper are determined by the physical conditions of the communication channel 
and cannot be controlled. 
 Therefore, the security relies on Wyner wire tap coding or coset coding (in some cases, privacy amplification). 
 This difference between the two systems is one of the reasons why QKD researchers misunderstand this issue.

\begin{figure}
\centering{\includegraphics[width=8cm]{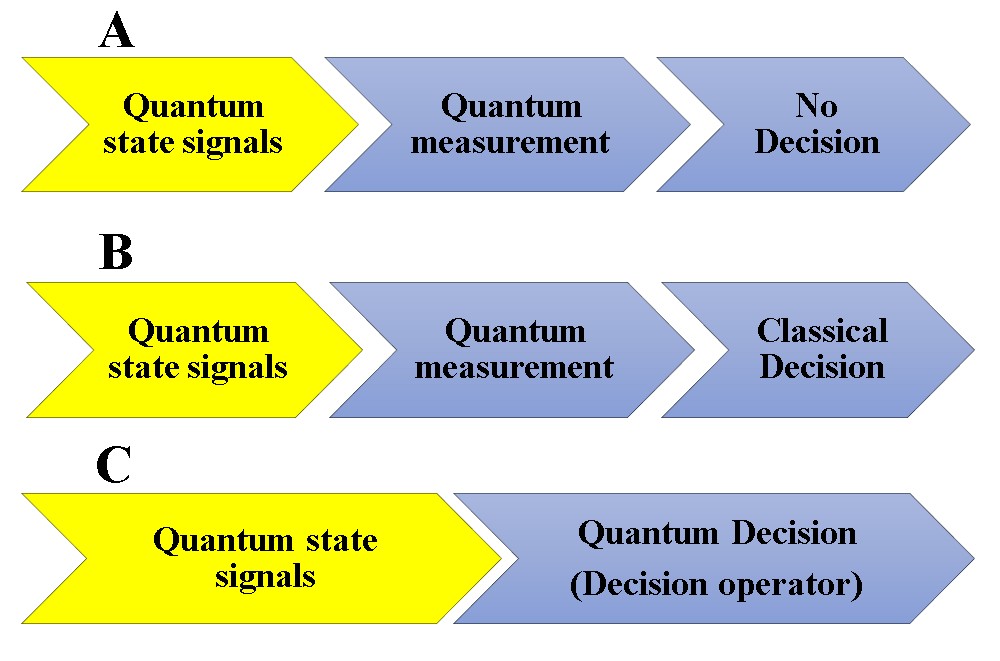}}
\caption{\textbf {Comparison of the quantum measurement and quantum decision described by POVM. }
A shows the simple quantum measurement for Born effect. 
B shows the application of classical communication theory for decision. 
C shows the quantum decision by decision operator derived from quantum communication theory. 
The asymmetry of channels of Bob  with key and 
Eve  without key come from the decision in the cases of B and C.}
\end{figure}

\section{\textbf{Mathematical structure of encryption and decryption of quantum stream cipher}}
\subsection{\textbf {Structure of quantum signals}}

As a first step, a set of communication basis of $M$ for transmitting binary data (plaintext) is defined as follows.

\begin{equation}
{\bar \theta_m}=\{|\alpha e^{i\theta_m}> , \quad |\alpha e^{i(\theta_m +\pi)} >\}
\end{equation}
where $|\alpha e^{i\theta} >$ is the quantum state vector called the coherent state of laser light, $\alpha$ and $\theta$ 
correspond to  amplitude and phase of light wave.
A communication basis is selected by running key $m=1,2,3,\dots, M \in K^R$ generated from PRNG with secret key $K$.
Then data (plaintext) is transmited by the selected basis as follows:
\begin{eqnarray}
X=0\rightarrow \theta_m,&& X=1\rightarrow \theta_m+\pi, \quad m=even \nonumber \\
X=1\rightarrow \theta_m,&& X=0\rightarrow \theta_m+\pi,\quad  m=odd
\end{eqnarray}
This is called encryption by modulation.
As a result, the phase signal has the following structure.
\begin{equation}
\theta_j=f(X, K^R)
\end{equation}
where $f(\cdot)$ is a mapping from electric signals to optical signals, $X=0,1$, $K^R=1,2,3,\dots, M$, and $j=1,2,3,\dots,2M$. 
Thus, the total number of optical phase signals is $2M$.
Such an encryption function is realized by several optical quantum modulation systems.

In this case, a set $\{\theta_j\}$ corresponds to the ciphertext before the mesurement. They are transmitted as following $2M$ quantum coherent states.
\begin{eqnarray}
&&|\alpha e^{i\theta_1} >, |\alpha e^{i(\theta_1+\Delta)} >, \dots,  |\alpha e^{i(\theta_1+M\Delta)} >,\\
&&|\alpha e^{i(\theta_1+\pi)} >, |\alpha e^{i(\theta_1+\pi+\Delta)} >, \dots,  |\alpha e^{i(\theta_1+\pi+M\Delta)} >\nonumber
\end{eqnarray}
where $\Delta=\pi/M, M \gg 1$.

Alice and Bob have the same secret key and PRNG, Bob can use the phase control device to convert back to binary state signals by using a running key 
from the same PRNG to the $2M$ state signals randomized by Alice's PRNG. The quantum detection model becomes the binary quantum states 
independent of the communication basis.

So Bob's decision becomes as follows (See Fig.2, Fig.3):
\begin{equation}
|\alpha > \quad  or  \quad |\alpha e^{i\pi} > \quad \forall m \in M
\end{equation}
This corresponds to the reception of amplitude $\alpha$ or $-\alpha$ under $\alpha \gg 1$. 

Thus, Bob can receive the data directly from the receiver using the key information.
Then, the average error probability on data $X$ is given by Helstrom receiver or the homodyne receiver as follows [13]:
\begin{eqnarray}
{\bar {P}_e^B}(X)&=&\min_{\{\Pi\}}\{1-\sum_{m=0}^{1} \xi_m Tr \rho_m^B \Pi_m \} \nonumber \\
&=&\frac{1}{2}[1- \sqrt {1-4\xi(1-\xi) Tr(\rho^B_0 \rho^B_1)}] \nonumber \\
&<&  P_e(Homodyne)  \ll \frac{1}{2}
\end{eqnarray}
where $P_e(Homodyne)$ is the error performance when Bob adopts the homoyne receiver for binary PSK.
So Bob can receive with almost no errors. Hence we have 
\begin{equation}
H(X|Y^B(K)) \cong 0
\end{equation}
where $Y^B(K)$ is the received signal by using phase control with key.

\begin{figure}
\centering{\includegraphics[width=8cm]{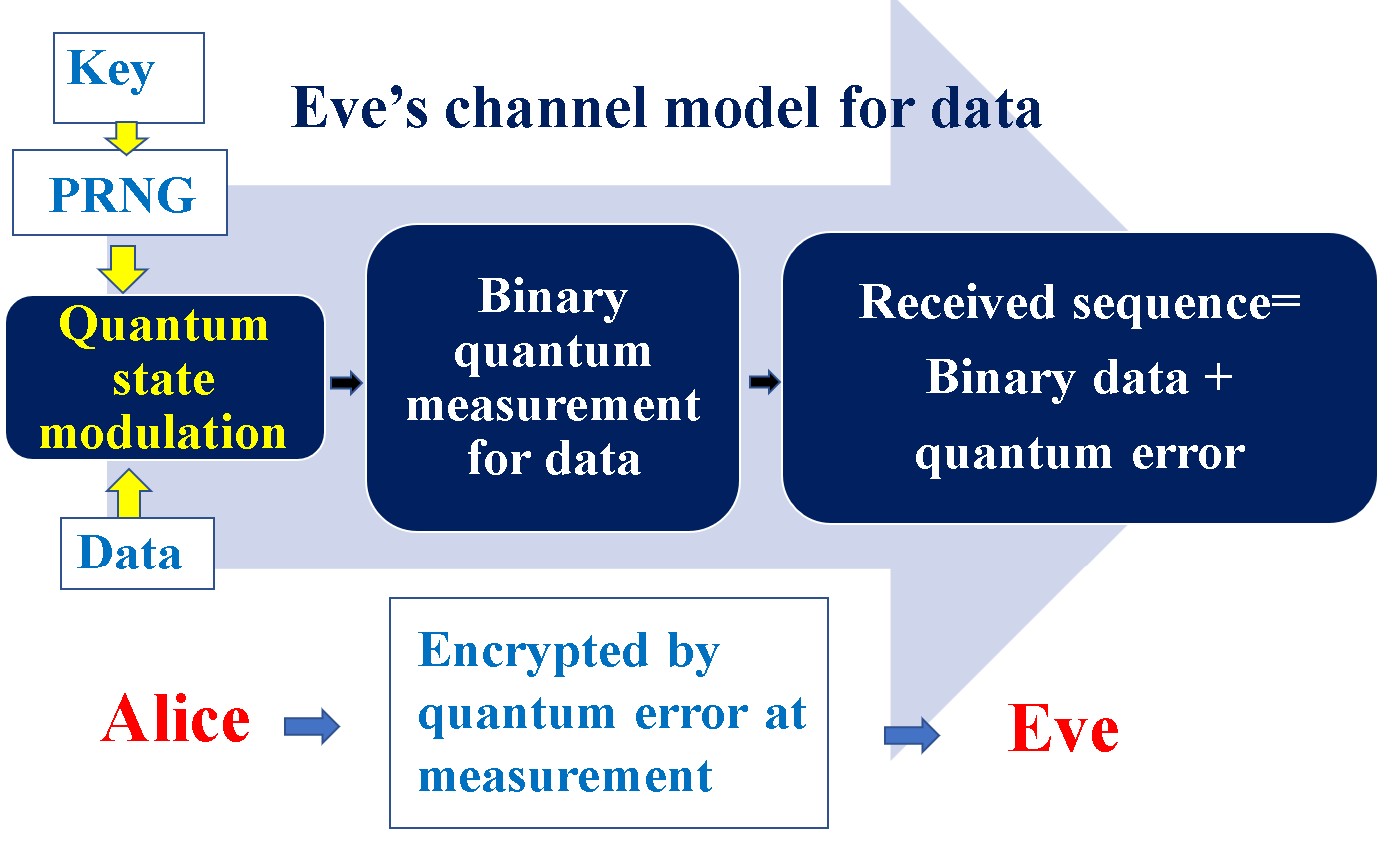}}
\caption{\textbf {This shows a channel model of attack against data.} 
Data sequences obtained by an eavesdropper in quantum stream cipher is encrypted 
at measurement by quantum error. The quantum error can be controlled by Alice's randomization based on modulation scheme.
 In the case of conventional cipher, received sequence is ciphertext enrypted by PRNG.}
\end{figure}

\begin{figure}
\centering{\includegraphics[width=8cm]{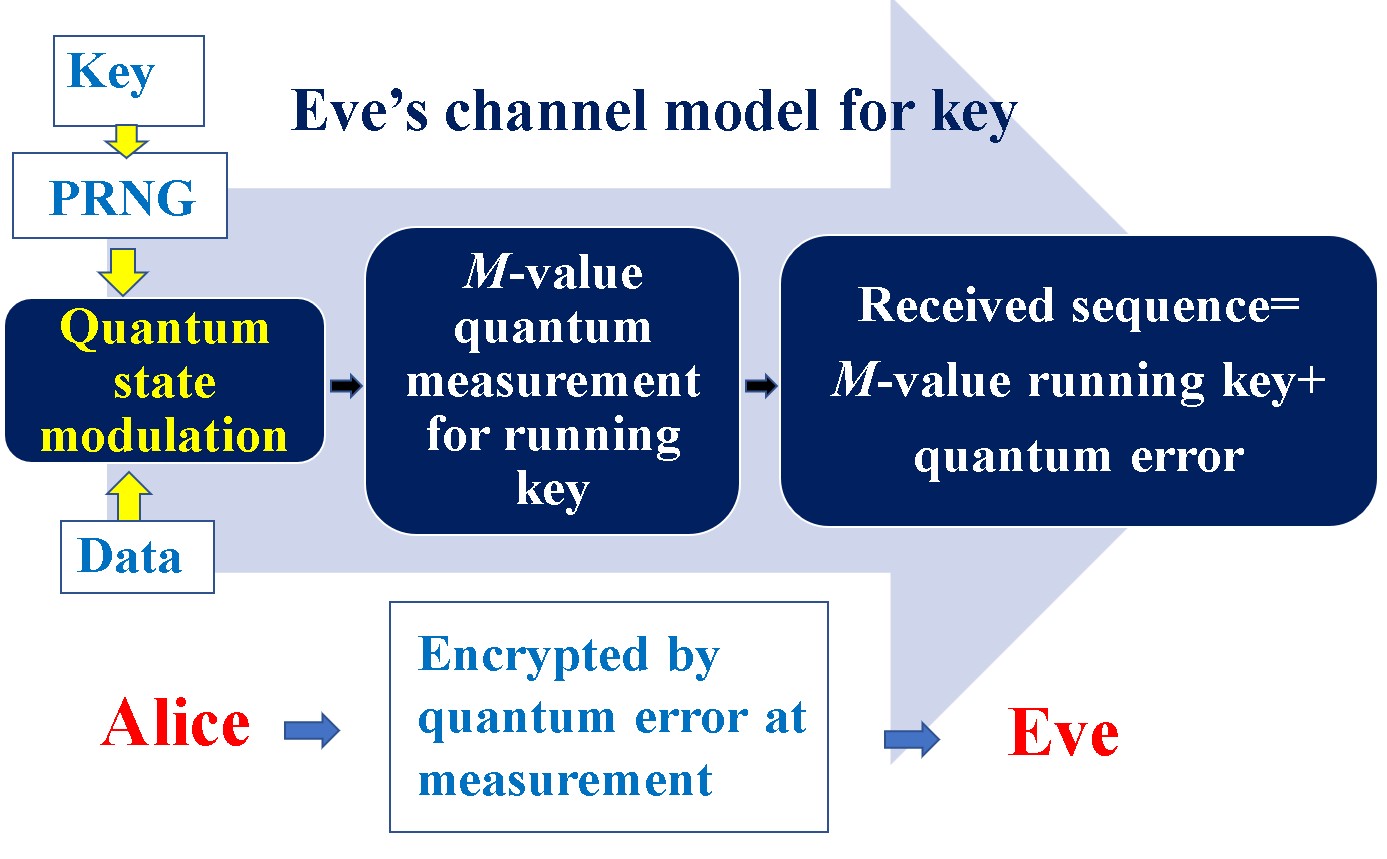}}
\caption{\textbf {This shows a channel model of  attack against key. } The received quantum state signal before the mesearment is the same as in FIG. 5, 
but the reception control system must change depending on what information is extracted. 
In the case of KPA,  the received signals are regarded as $M$-value running key sequences, the receiver must be designed to $M$ value signals. 
The decision values  obtained by an eavesdropper  are equivalent of ciphertext encrypted by quantum error. In the case of conventional cipher, 
received sequence is running key itself.}
\end{figure}

\subsection{\textbf{Eve's receiver designe to obtain data and key information}}
Although the signal  before the receiver is in the $2M$-valued signal format, the receiving system to be adopted must be different 
when receiving data and when receiving key information. In this section, we will explain the reason.

\subsubsection{\textbf {Data}}
Assuming that an eavesdropper analyzes information about the data, the model as encryption mechanism is as shown in Fig.5, 
because of the signal structure (Fig.3).
When  Eve adopts the heterodyne receiver, the quantum measurement process corresponds to B-type in Fig.4.
Then, she needs to distinguish between two adjacent coherent state signals:
$|\alpha e^{i\theta_1} >, |\alpha e^{i(\theta_1+\Delta)} >$ with signal distance $\Delta$ in Eq(21). 
When $M \gg 1$, the error becomes $\frac{1}{2}$. This corresponds to B-type in Fig.4.

On the other hand, if Eve  would adopt the binary quantum optimal measurement for the following mixed quantum states, 
the quantum measurement process corresponds to C-type in Fig.4.
\begin{eqnarray}
\rho^E_0 &=&\frac{1}{M}\sum_{m=1}^{M} |\alpha_{(m=even)}><\alpha_{(m=even)}| \nonumber \\
\rho^E_1 &=&\frac{1}{M}\sum_{m=1}^{M} |\alpha_{(m=odd)}><\alpha_{(m=odd)}| 
\end{eqnarray}
This structure of mixed state is called doubly symmetric mixed state, and the quantum Bayes(also minimax) solution for 
 such as mixed states of coherent state was given in [14].
Then the average error probability for binary data  in the case of $M\gg 1$ is given as follows: 
\begin{equation}
{\bar {P}_e^E}(X)=\max_{\{\xi\}}\min_{\{\Pi\}}\{1-\frac{1}{2}\sum_{l=0}^{1} Tr \rho_l^E \Pi_l \}\sim \frac{1}{2}
\end{equation}

\subsubsection{\textbf {Key information}}
In order for Eve to perform the cryptanalysis for key, she has to obtain the information of 
running key sequence by her quantum measurement to $2M$-value (or M-value for known plaintext) quantum ciphertext.
The first step in the procedure leading to an attack is to receive a signal flowing through the real communication channel. 
Thus, the encryption mechanism on running key sequence is given as Fig.6, because of the signal structure.

Let us assume that an eavesdropper adopts the receiving system of type B in Fig.4.
It corresponds to a heterodyne receiver to detect one of the $2M$ or $M$ signals within a rough signal reception range,  and 
the detection probability would be approximately as follows, depending on the number of signals $\Gamma$ in the region of 
the quantum noise masking effect:
\begin{equation}
P_d(Y^A) \sim \frac{1}{\Gamma}
\end{equation}
where $Y^A$ corresponds to the running key  $K^R$ of $M$ value of the output from PRNG.

If she adopts the quantum receiver, the average minimum error probability for the adopted quantum state signal scheme
 (or equivalently maximum detection probability) can be given by the formulae of quantum detection theory [12,15,16,17]. 
That is, the Eve's error performance of the information on the running key  in this situation  is given by 
\begin{equation}
{\bar {P}^E_e}(Y^A)=\max_{\{\xi\}} \min_{\{\Pi\}}\{1-\sum_{m=1}^{2M} \xi_m Tr \rho_m \Pi_m\}
\end{equation}

Fot the indivisual and collective schemes, the optimum solution of the above foumula for coherent state signals is given in the reference [18].
These provides the evaluation of the theoretical accuracy of the running key that the eavesdropper can obtain.

\subsection{\textbf { Mechanistic explanation of security}}

As explained above, the security of quantum stream cipher based on Y-00 protocol is evaluated by assessing how accurately an eavesdropper 
can receive the signal. Below, we explain how this is done.\\

\subsubsection {\textbf {Mechanism for ensuring data security }}

When considering data security, it is important to understand that the transmitted $2M$-valued signal is treated as a signal carrying plaintext information.
In this way, when an eavesdropper receives a signal using a receiver without a key, the plaintext is encrypted by quantum error
generated during binary decision at measurement, corresponding to Fig.5.

There are two ways to receive it. 
One is to use heterodyne scheme, which allows adjacent signals among the $2M$ signals to be distinguished. 
However, when $M \gg 1$, the distance between the adjacent signals becomes extremely small, falling into the quantum noise mask region 
and making discrimination impossible. In other words, the error probability is 1/2. This is equivalent to the plaintext sequence being masked 
by a completely random number.

Another reception method is quantum optimal detection scheme to a mixed state corresponding to 0 and 1 based on plaintext (data) statistics.
 According to numerical analysis of the convergence of the error probability in this case,  a significant number of  $M$ is required for perfect security.\\

\subsubsection {\textbf {Role of overlap selection keying:OSK }}

In 2003, Sohma et al. adopted a method of additional randomization to improve the above properties [19]. 
Let us explain it in the following.  
A  physical phase signal on the phase plane simultaneously carries both plaintext and basis information (running key). 
Adjacent signals alternate between 0 and 1. 

Then, a new randomization is introduced to these 0 and 1 using PRNG-2 of the same key as the selection of communication basis. 
Fig.7 shows the scheme. Such scheme is called OSK (Overlap selection keying). 

This makes it even more difficult to distinguish the plaintext, because the density operator for 0 and 1 of data becomes the same for 0 and 1.  That is, 
\begin{equation}
\rho^E_0 = \rho^E_1 \longrightarrow \bar{P}^E(X_{Y-00}) = \frac{1}{2}
\end{equation}
Thus, they become indistinguishable, resulting in a perfect error. 
In addition, the plaintext is automatically converted into encrypted plaintext. It is called Y-00 plaintext.
\begin{equation}
X \longrightarrow X_{Y-00}(X, K)
\end{equation}
But Bob can receive the correct plaintext $X_{Y-00}$ by correct running key information and decrypt $X_{Y-00}$ by the same key.

\begin{figure}
\centering{\includegraphics[width=8cm]{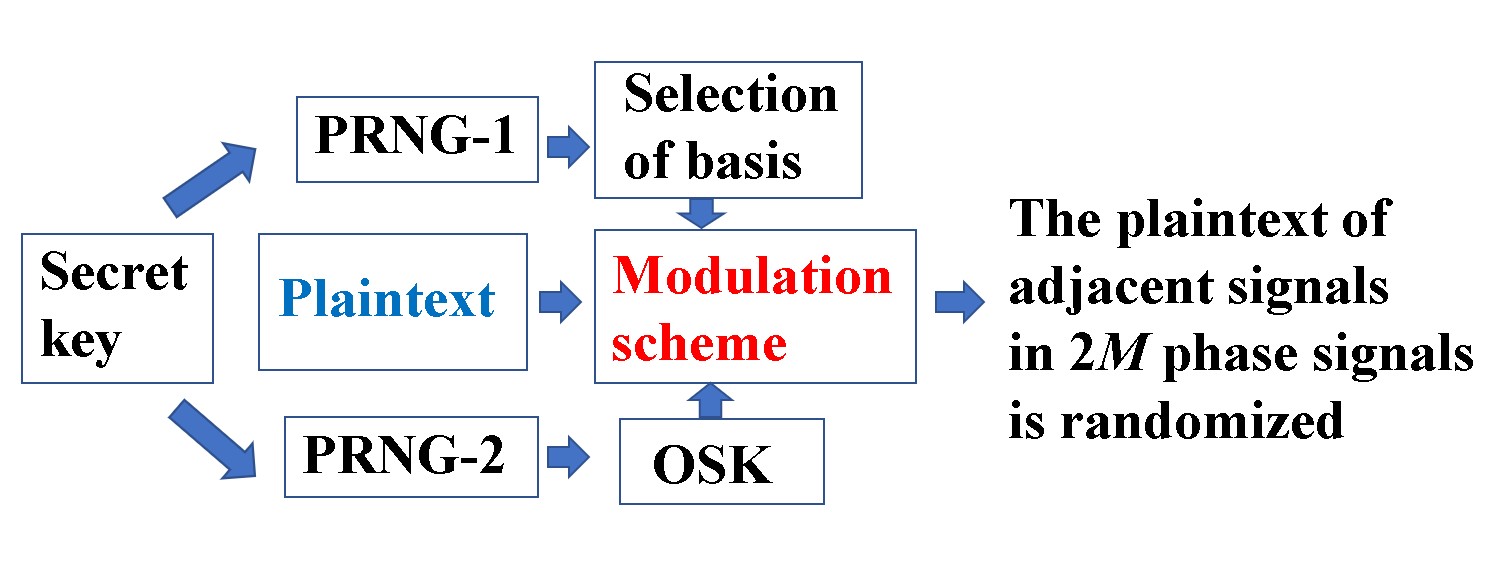}}
\caption{\textbf {The scheme of OSK.}  Each phase signal on the phase plane in Fig.2 is assigned by plaintext and basis information. 
Adjacent signals are assigned alternately as 0 and 1. OSK is equivalent to randomizing the alternate assignment..}
\end{figure}

\begin{figure}
\centering{\includegraphics[width=8cm]{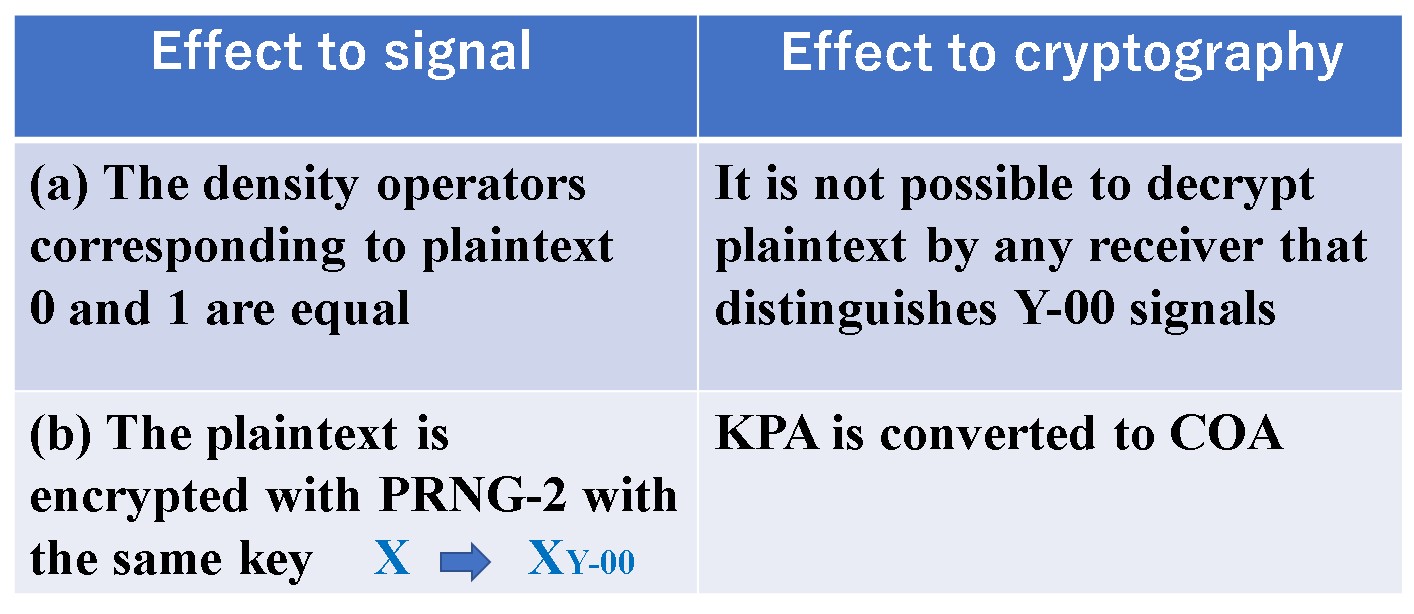}}
\caption{\textbf {The effects of OSK.} KPA is known plaintext attack, and COA is ciphertext only attack.}
\end{figure}

As a result, despite the use of a running key in which a short key is mathematically extended, complete information-theoretic security 
can be guaranteed against ciphertext-only attacks on plaintext (data) with arbitrary statistical structure.

In this way, this scheme is no longer bound by the Shannon impossibility theorem. 
At this point, the criticism that ``Y-00 is the same as conventional encryption" is invalid.
 
Let us explain another role of OSK. This effect was not known when OSK was proposed. In the model without OSK,  
the $2M$ phase signals are uniquely assigned by the basis information and plaintext. When OSK is performed, even with the same phase signal, 
the plaintext can be either 0 or 1. Therefore, the relationship between the plaintext and the basis information for the $2M$-value phase signal 
becomes non-injective. This means that even if the original plaintext is known, the $2M$-value phase signal corresponding to the ciphertext can be 
regarded as the basis information (running key), but since there is no correspondence with the plaintext, even if an eavesdropper attempts 
a known-plaintext attack, it becomes ciphertext only attack.  Such properties are summarized in Fig.8.\\

\subsubsection{\textbf {Mechanism for ensuring key security}}
The next most important factor in the security of symmetric key cryptography is the security of the key. Key information is equivalent to analyzing 
the running key. Because the basis is directly selected by the running key, we must consider the receiving system that receives 
the communication basis information. 
In other words, because the structure of the signal gives that the $M$-value phase signals directly conveys the basis information, 
Eve adopts a receiving system for the $M$-value phase signals. This allows to adopt  heterodyne receiver with $M$-value decision scheme 
or quantum optimal receiver.

Here, let us consider exhaustive search as an attack against the key in Y-00 protocol. Exhaustive search is a method in which, 
when the plaintext statistics or other information are completely known, the correct key is determined by trying all possible keys and 
comparing them with prior information. In mathematical cipher, the correct detection probability is 
\begin{equation}
P_d(K|X_n, Y_n) = 2^{-(|K|-n)}, \quad \forall  n \le |K|
\end{equation}
If the known plaintext is the key length ($n=|K|$), an exhaustive search of keys provides $P(K|X_{|K|}, Y_{|K|})=1$  in Eq(31)
 because of the structure of the conventional mathematical cipher. 

Let us consider the case of Y-00 protocol.  In heterodyne reception, 
the number of signals on the phase plane that are masked by quantum noise is defined as $\Gamma < M$ (See Eq(27). 
The receiver output value, including any errors, is stored in memory. Let $N$ be  $N=|K|/log M$. 
Next, assume there are $N$ known plaintexts. 
An exhaustive search can now be attempted.
When the plaintext information is arranged alternately in a $M$-value phase signal, if the plaintext is known, 
the measurement value will always have an error within the range of $\Gamma /2$. Therefore, the ambiguity of a key is expressed as 
a probability as follows:
\begin{equation}
P_d(K|X_n, Y_n) \cong  (\Gamma/2)^{-{\frac{|K|}{log M}}}, \quad  \forall n \le N
\end{equation}

This probability does not converge to 1 even if the number of known plaintexts increases.
\begin{equation}
P_d(K|X_n, Y_n) \ne 1,  \quad  \forall n \ge |K|
\end{equation}
This is because once the running key exceeds the key length, the running key sequence becomes a deterministic sequence, 
and further observations may not provide 
any information about the key.

Now, consider the case where OSK is introduced. In this case, the  known-plaintext attack does not work.
Therefore, the probability of detecting the key is a simple guessing probability as follows:
\begin{equation}
P_d(K|X_n, Y_n) = (\Gamma)^{-{\frac{|K|}{log M}}}, \quad \forall n
\end{equation}

Next, let us consider the attack based on mathematical analysis of the measured sequence.  
Given the structure of the Y-00 protocol, it is possible to treat the signal on the phase plane as only the communication basis information (running key information) as shown in Fig.6. 
This configuration allows for direct analysis of the output of PRNG using a receiver that identifies $M$-value signals. 
In the case of mathematical cipher, Eve's received sequence is only the running key sequence from the PRNG in the case of KPA, or a running key sequence with randomness due to plaintext in the case of COA. 

On the other hand, in Y-00, the received sequence is a running key sequence randomized by quantum errors. 
Thus, the uncertainty of the key is a two-fold structure consisting of the PRNG and quantum errors in the quantum measurement.
An uncertainty region due to quantum error is $\Gamma/2$ ($\Gamma$ in the case of  OSK) .

Then, the information-theoretic evaluation is performed using the generalized unicity distance. 
The details of this theory are explained in the next section.

\section{\textbf {Quantitative evaluation of key security}}

\subsection{\textbf {Theory of generalized unicity distance}}
Quantitative evaluation of the security of quantum stream cipher may be carried out directly by the error characteristics
 of the eavesdropper's ciphertext, but its formulation is still a subject of consideration.
 
Thus, to develop the theory of generalized unicity distance is  preferable to evaluate the performance of information theoretics security.
Even so,  due to the complexity of the model, it is difficult to construct a theory to determine the exact unicity point 
which gives the correct key with probability 1.
So a lower bound method is used.  It is given as follows [9, 12]:\\

$\textbf {Theorem 3}$ $\{Yuen \cdot Nair\}$\\
The lower bounds of the generalized unicity distance for COA and KPA are given as follows:
\begin{eqnarray}
n^Q_0 &>& \frac{|K|}{C_1(COA)},\\
C_1(COA)&=&\max_{\{\Pi^E\}}I_{COA}(K^R;Y^{E_q}) \nonumber \\
n^Q_1 &>& \frac{|K|}{C_1(KPA)},\\
\quad C_1(KPA)&=&\max_{\{\Pi^E\}}I_{KPA}(K^R;Y^{E_q})\nonumber
\end{eqnarray}
where  $C_1$ is the maximum amount of the mutual information by the eavesdropper's measurement from set of quantum states with running key information, 
which is called an accessible information.  $C_1$ can be replaced by Holevo capacity $\chi_H$ 
based on the collective measurement attack which is taken into account as the upper bound for Eve's channel.\\

The theory of maximaization of mutual information is twinned with the quantum detection theory and 
has the same theoretical structure as the optimal theory. It is derived by Holevo as follows [15]:\\

$\textbf {Theorem 4}$\{Holevo\}\\
The necessary condition for maximum mutual information with respect to the decision operators for a simple set of states 
is given as follows:
\begin{eqnarray}
&&P(j|i) =Tr \rho_i \Pi_j \nonumber  \\
&& {\bf F}_j = \sum_{l} \xi_k \rho_k \log  \{\frac {P(j|l)}{\sum_k \xi_k P(j|k)}\} \nonumber \\
&& \Pi_j[{\bf F}_j - {\bf F}_i]\Pi_i =0,  \forall i,j 
\end{eqnarray}
\\
The method of approximating by an upper bound on the accesible information 
 is valid based on the following theorem [20,21].

$\textbf {Theorem 5}$\\
 $\{Holevo\cdot Schumacher \cdot Westmoreland \}$\\
The upper bound of maximum mutual information and the capacity are given by the Holevo information as follows:
\begin{eqnarray}
&&C_1 \le S( \rho_T) - \sum_{k=1}^{M} \xi_k S(\rho_k)=I_H\\
&&where \quad \rho_T = \sum_{k=1}^M \xi_k \rho_k, \nonumber\\
&&Then \quad \chi_H=\max_{\xi} I_H
\end{eqnarray}
where $S(\rho)$ is the von Neumann entropy, and $\chi_H$ is called Holevo capacity.\\

The greatest feature of the Y-00 protocol is that it is possible to control  Holevo capacity  Eve receives. 
 That is, $C_1$ and $\chi_H$ are the function of randomization by modulation :$\Xi$.  
 Then the main subject is to find a way to achieve the following  results :Eq(40) without the degradation of communication performance for Alice and Bob:
 \begin{equation}
 \min_{\{\Xi\}} C_1(\Xi) \longrightarrow 0\quad or \quad  \min_{\{\Xi\}}\chi_H(\Xi) \longrightarrow 0
 \end{equation}
Some examples are given in the appendix.\\

\subsection{\textbf {Summary of strucure and role of generalized unicity distance}}

\subsubsection{\textbf {Cryptographic model of quantum stream cipher}}

Because the concept of encryption in  quantum stream cipher based on Y-00 protocol is entirely new, it seems that misunderstandings 
arise at the start point of attack. To attack secret key of Y-00 protocol, a clear understanding of its cryptographic modeling is first necessary. 
Therefor, we here denote again a detailed explanation of the structure of Y-00 protocol.\\

(a) Role of Mathematical Cipher: The primary role of the mathematical cipher component is to control information-theoretic quantities 
such as the probability of error, SNR, and Holevo capacity at the receiver of an eavesdropper who does not possess the key, 
and to degrade these characteristics. In other words, it makes the communication channel characteristics 
between the legitimate  and the eavesdropper asymmetric.\\

(b) Why degradation of Eve's ability is possible:  Key information can only be obtained by useing receiver system for responding to 
$2M$-value (or $M$-value) quantum state signals. Quantum Shannon information theory states that the greater the number of signals, 
the greater the non-orthogonality and the worse the error characteristics.\\

(c) Cryptographic properties: When an eavesdropper receives a quantum state signal randomized by mathematical cipher, 
the output sequence from Eve's receiver is in the form of a c-number sequence the same as conventional cipher. 
In other words, attempting to read the plaintext or 
to read the running key sequence, received sequences are as follows (See Fig.5 and Fig.6):\\
\textbf{(i) }Output sequence of Eve's receiver for data = binary plaintext + quantum error (true random number)\\
(Contrast: In the case of mathematical cryptography, Eve's acquired sequence = binary plaintext + running key sequence generated by a PRNG).\\
\textbf{(ii) } Output sequence of Eve's receiver for running key = $M$-value running key sequence generated by 
PRNG + quantum error (true random numbers)\\
(Contrast: In the case of mathematical cryptography, Eve's acquired sequence = running key sequence generated by PRNG).

From the above, analyzing the security of Y-00 involves constructing a theory for quantitatively evaluating the security of plaintexts 
and keys for Eve's receiver output sequences (i) and (ii). At this stage, the focus of the analysis shifts to the statistical properties 
of the received sequence, and the information-theoretic framework becomes central rather than quantum mechanics itself.
 Therefore, it is sufficient to generalize conventional information-theoretic analysis 
 so that they can be applied to the above model. Its trial is called generalized unicity distance theory.\\

\subsubsection{\textbf {Role of generalized unicity distance theory}}
We explained that in order to evaluate the information-theoretic security of the key in Y-00 protocol, 
it is necessary to construct a unicity distance theory for a generalized random cipher model. As mentioned previously, 
the sequence received by an eavesdropper is a model consisting of running key sequences 
and quantum error by quantum noise. 
The unicity distance theory for such sequences is called the generalized unicity distance.
Here, we show the concept of Yuen-Nair unicity distance theory for generalized random ciphers [9, 12]. 

In conventional cryptography, when a ciphertext-only attack on a key based on a statistically structured plaintext, 
the unicity distance is calculated by dividing the key length by the redundancy of the plaintext. Thus, determining the unicity distance itself is
 the subject of discussion.
On the other hand, in a known-plaintext attack, the unicity distance is fixed at the key length, and the key is determined with probability one.

However, in Y-00 protocol, when evaluating the unicity distance, the formulation for the lower bound of the unicity distance is 
the subject of the evaluation, not to derive the formula of the unicity distance itself. 
That is, the lower bound is the parameter of control by a mechanism between legitimate communicators. 
Then it is set to evaluate the ability to make the lower bound very large. The final evaluation of key is the probability of determining the key. 
Its formula will be reported in sebsequent paper.
Thus, even if the theoretical schemes of both conventional and generalized version seem to be the same, the operational meaning is completely 
different.

Previous criticism of the security of keys has been based on the argument that the lower bound of the unicity distance is the unicity distance point.
In that case, the unicity distance point is regarded as  the number of known plaintext :$|K|/log M$. 
At first glance, $|K|/log M$ is shorter than $|K|$, which gives the illusion that it is worse than conventional encryption.  
Such interpretation assumes the operational meaning of the unicity distance in the conventional theory. 

It is important to understand that even if the unicity distance theory has a similar format, its operational meaning is different. 
That is, in Y-00 protocol, there does not exist the unicity distance point which can determine the key with the probability one.

This is because, given that the original key length is $|K|$, and subsequent running keys are pseudorandom and it becomes deterministic. 
The observed sequence is equivalent that completely random numbers are superimposed on its running key sequence, no matter how long the observations, the eavesdropper's information will not increase. The theoretical back ground on this issue is given in the reference  [22], the section 4 in the reference [23], and  [24]. 
Thus, the purpose of generalized unicity distance theory is to control the lower bound, and to find a way of $n^Q_1 \gg |K|$.\\

\section {\textbf {Conclusion}}
In this paper, we explained the mechanisms for ensuring the security of various cryptographic technologies.
We have emphasized that Y-00 protocol differs from conventional cryptographic mechanisms in that it incorporates a mechanism 
that amplifies the effect of quantum noise on eavesdroppers during the operation process of the communication system. 
So we explained how this mechanism encrypts transmitted information, etc., in comparison with other cryptographic technologies. 

According to the above results, the security against ciphertext only attack on data (plaintext) is the same information theoretic security as 
a one time pad, despite the short secret key scheme. 
 
When it comes to the information theoretic security of secret keys, discussions are a bit more complicated. 
In general, the quantum noise effect of the standard quantum stream cipher scheme is small. 
Its fact gives that  the lower bound of unicity distance  is not very large. But it does not mean that the security is broken, 
it means that the lower bound is small.  
The correct detection probability of the key is still not 1. This performance  cannot be realized by the conventional cipher.

Here, one should keep in mind that the quantum stream cipher may be able to guarantee complete security for the key, 
as there is room to add features that increase the eavesdropper's error probability or reduce the Holevo capacity. 
Its subject is called a generalized quantum stream cipher. 
The advanced studies towards the realization of the generalized quantum stream cipher are given in [10, 11, 25].

In order to provide a concise explanation of the theoretical flow, many contributions have been omitted.
It is our sincere hope that this commentary will contribute to mutual understanding in each field.

 \section*{\textbf{Acknowledgements}}
We are grateful to the academic members of the group seminar held in Nov. 2025. This lecture note is the extended  version of  the note in the seminar. 
We also  thanks  Marketing Research Group  for the discussions and their activities.
 
\section*{\textbf{Appendix}}
In this appendix, we provide a summary of the comparison of the security guarantee principles of three representative cryptographic methods for the information theoretics security.

\subsection {\textbf {Security mechanism of physical layer security  for microwave wireless scheme}}
The physicsl layer security  in wireless systems basically uses Wyner's wiretap channel scheme. Hence, the 
information-theoretically secure secret communication or key distribution in its enviroment is constrained by the necessary condition 
that the SNR of an eavesdropper must be degraded. 
Under the above condition, it is able to design a code that an eavesdropper cannot decode correctly, based on the concept of 
the Shannon channel coding theorem.
These are called wire-tap codes or coset codes [26, 27]. 

The most important issue as the first step is to develop a technique to degrade the SNR of the eavesdropper compared to the SNR 
of the legitimate communicator. 
In a wireless communication environment, such conditions may be relatively easy to achieve.
Currently, the following technologies are being developed:\\
(a) Beamforming\\
(b) Artificial Noise\\
(c) Cooperative Relaying\\
(d) Reconfigurable Intelligent Surface\\
Then, following Wyner's theory, the secret capacity is calculated, and the following codes are being considered for their implementation and 
construction such as polar codes or LDPC. (See [28],[29]])

In light of this situation, one can consider a key distribution technology by using Wyner scheme:\\

$\quad $ Key disribution =PLS + privacy amplification\\

Then the security metrics is given by Pinsker inequality as follows:
\begin{equation}
d_c \le \sqrt {\frac{1}{2} I(K; Y^E)}
\end{equation}
where $d_c $ is statistical distance. 
A more sophisticated method based on information spectrum theory will be presented in the subsequent lecture note.

\subsection{\textbf{Security mechanism of QKD}}
 Let us denote concepts behind security assurance techniques for quantum key distribution. Quantum key distribution theory begins with 
 Wyner's theory of eavesdropping channels. It is QKD-a in this paper. Generally, the Wyner scheme requires that the SNR of 
 the eavesdropper's channel be worse than  the SNR of the legitimate channel. To realize the equivalent condition of this, 
  QKD employs classical communication between legitimate communicators in addition  to quantum communication channels. 
  Therefore, security must be guaranteed for quantum, classical, and hybrid systems.

Early theories considered the effect of quantum disturbance in the signal interaction of quantum communication channels, calculated 
the secrecy capacity $C_{se}$  based on this effect, and designed coding systems under 
the conditions of that rate. 

However, it was pointed out that one needs to ensure security for hybrid systems that also incorporate classical communication channels,
 quantum collective attacks.
It is composed  by storing qubit strings in quantum memory and incorporating information from the classical system.

According to Devetak-Winter theorem [4], the maximum mutual information of the eavesdropper's channel in the collective attack is the Holevo capacity, 
 and the secret capacity is 
\begin{equation}
C_{sc} =I(A;B) - \chi_H(X;Y^E)
\end{equation} 
Thus, in early QKD security proofs, Holevo capacity played a central role in determining the Wyner secret capacity.

Meanwhile, it was pointed out that because the sequence transmitted by QKD is a secret key for symmetric key cipher, 
the sequence shared between Alice and Bob must appear completely random to an eavesdropper. 
Therefore, the evaluation criterion was converted to the $\textbf {trace distance}$ :$d_q$ between ideal
 $\rho_{K}\otimes \rho_E$ and real $ \rho_{KE} $ density operators.
\begin{equation}
d_q=\frac{1}{2}|| \rho_{KE} - \rho_{K}\otimes \rho_E||
\end{equation}

Then, the metrics for the security  was converted to  min-entropy $H_{min}$ as follows. 
\begin{equation}
H_{min}(X|Y^E)= -\log \max_x P(X=x|Y^E) 
\end{equation}
where $X$ is plaintext and $Y^E$ is received sequence of Eve.

This scheme is denoted as QKD-b in this paper.
At this time, the generation of the sequence of uniform randomness within the  min-entropy range is guaranteed by
the $\textbf {quantum leftover hash lemma}$ [5]. The technique to atain it is the privacy amplification.

An upper bound on the length of the sequence to be provided for the symmetric key cipher is assured by 
the smooth min-entropy under the trace disatnce $d_q \le \epsilon$. 

In what follow, we denote its theoretical process. Min-entropy is a quantity that is constrained by physical conditions of the communication channel. 
This quantity ultimately determines the final secure sequence length. Therefore, it is important to estimate this quantity. 
In this case, one can use the Holevo capacity to evaluate min-entropy in the asymptotic limit. That is, the lower bound of min entropy is given by
\begin{equation}
H_{min}(X|Y^E) \ge H(X) - \chi_H(X;Y^E)
\end{equation}
Here, the internal variable of the above functional is  the physical parameters of the communication channel.

Thus, in current composable security scheme, Holevo capacity becomes an auxiliary quantity used to provide a lower bound on  
min-entropy in the asymptotic limit.

In sum, the security of QKD can be summarized as follows:
Based on min entropy, which is determined by the parameters of the communication channel, and the value: $\epsilon $ ($d_q \le \epsilon $) 
of the trace distance set by the designer, 
it is guaranteed that the sequence shared by legitimate communicators will appear to an eavesdropper as a truly random sequence within the range $\epsilon$.
 The technique for achieving this becomes privacy amplification.
 Finally, the length of the sequence for which security is guaranteed is determined by smooth min entropy.

For a detailed history and the theory of finite analysis, see the literature [4, 5, 30, 31, 32, 33].

\subsection{\textbf{From standard quantum stream cipher to generalized quantum stream cipher}}

The information-theoretic security of quantum stream ciphers using the Y-00 protocol originates from the mechanism that randomizes 
the modulation mechanism between legitimate communicators, thereby causing large errors in the decision of signal of an eavesdropper 
who does not possess the key. 

That is,  the quantum stream cipher is a method that artificially degrades the noise characteristics of receiver of an eavesdropper 
by randomization of modulation scheme. 

The metrics of security on secret key becomes the generalized unicity distance, 
because Eve's analysis becomes analysis of sequence disturbed by  quantum error.

As we pointed out, ``the lower bound of the unicity distance" is calculated by dividing the key length by the access information of Alice-Eve 
channel or the Holevo capacity which is the upper bound of it as follows:

\begin{equation}
n^Q > \frac{ |K|}{ \chi_H(Y^A;Y^{E_q}|\Xi) }
\end{equation}
where $ \chi_H(Y^A;Y^{E_q}|\Xi)$ is the Holevo capacity of the channel from the running key ($Y^A$) of Alice to 
the observation sequency of Eve :$Y^{E_q}$ and it is controllable with respect to the randomization by modulation $\{\Xi\}$.
This feature is derived from the mechanism in the Y-00 protocol that allows control of communication-theoretic characteristics related to 
receiver performance, as shown in Fig. 9 and Fig.10.

In general, because the quantum noise masking effect $\Gamma$ of the standard quantum stream cipher is significantly smaller than $M$, 
the Holevo capacity cannot currently be made negligible. Therefore,  information-theoretic security for the key is not complete. 
From the above, it can be concluded that standard quantum stream ciphers are information-theoretically secure for arbitrary plaintexts, 
even with short keys, but information-theoretic security for the key still has issues.

In general, the standard quantum stream ciphers serve as an explanation of the principle of Y-00 protocol. The above performance is not final one. 
That is,  this cryptographic mechanism does not deny the ability to guarantee the information-theoretic security of the key. 
In other words, the next challenge is to develop randomization techniques to achieve complete information-theoretic security for the key. 
Pioneering research on this topic has been proposed in the literature, so it is expected that many researchers will build on this foundation 
and take on the challenge.
As example, we show  two methods as follows: [10, 34].\\
(a) Randomization method by product cipher form. \\
Delibelate signal randomization (DSR), and Quantum noise diffusion mapping (QNDM).\\
(b) $M$-th order extended quantum code modulation. \\
Coherent PPM, and FM-phase PPM.\\

 \begin{figure}
\centering{\includegraphics[width=8cm]{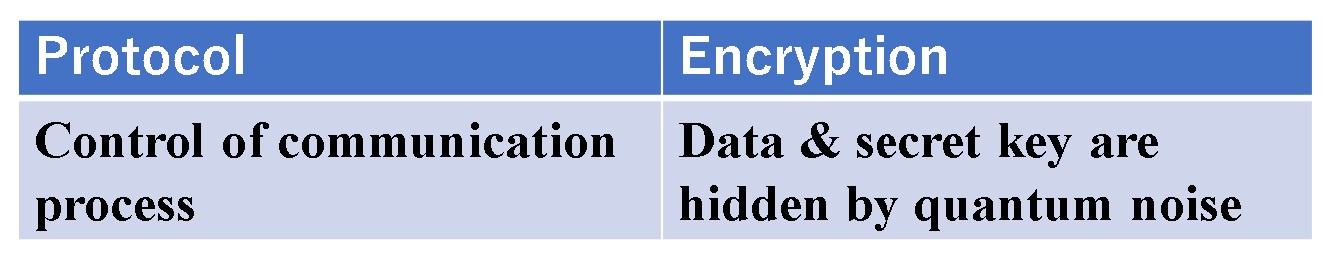}}
\caption{\textbf {Encryption mechanism of Y-00 protocol: A.} 
Cryptographic functions are based on concepts from communication theory.}
\end{figure}

 \begin{figure}
\centering{\includegraphics[width=8cm]{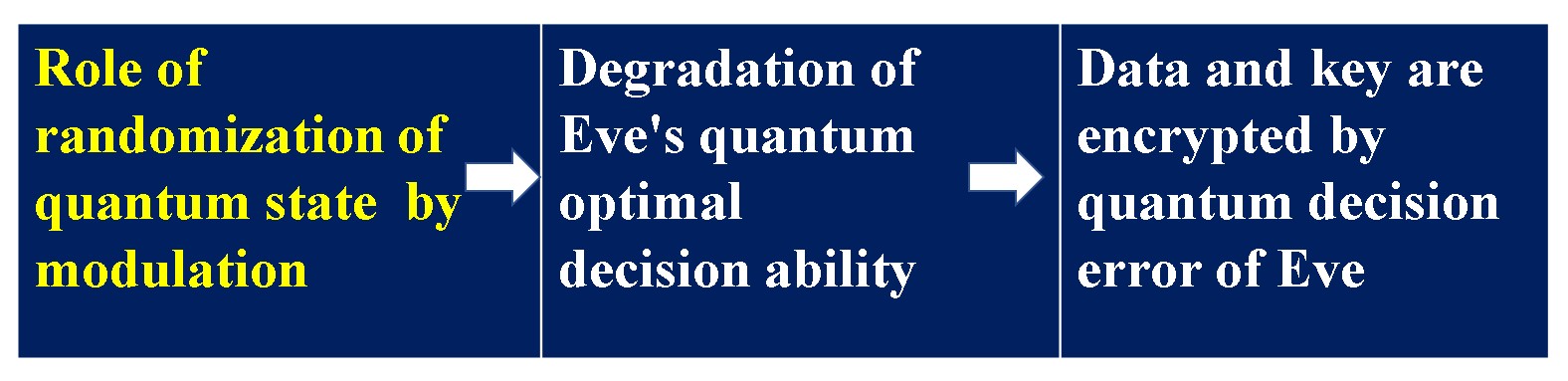}}
\caption{\textbf {Encryption mechanism of Y-00 protocol: B.} Encryption is done when Eve measures signal and make decision based on 
the measurement. The error performance of Eve is controlled by randomization of quantum state by modulation schemes. }
\end{figure}

\subsection{\textbf{Comparison of PLS, QKD, and Y-00}}
Finally, we will compare the differences in the requirements for evaluating the security mechanisms of PLS, QKD and Y-00 protocol. 
As mentioned above, PLS and QKD require information about the communication channel used to calculate security metrics, 
but Y-00 does not as shown in Fig. 11.

Meanwhile, the subjects of attack against these techniques are described in comparison with Fig.12.  
In the case of PLS in the microwave domain, it does not need the quantum attack. 
In the cases of QKD and Y-00 protocol, the quantum attacks are main issues, but the methods for both schemes are quite different.

In order to clarify it, we denote a comparison of the physical phenomena that guarantee security between QKD and Y-00  protocol 
in Fig.13.  QKD uses the physical phenomenon itself, but Y-00 protocol uses the communication theoretic performance based on a physical phenomena.

Since these are physical cryptography, device development is required.
In this way, quantum stream cipher is  going to develop in a similar manner to the development of QKD, as shown in Fig.14.

\begin{figure}
\centering{\includegraphics[width=8cm]{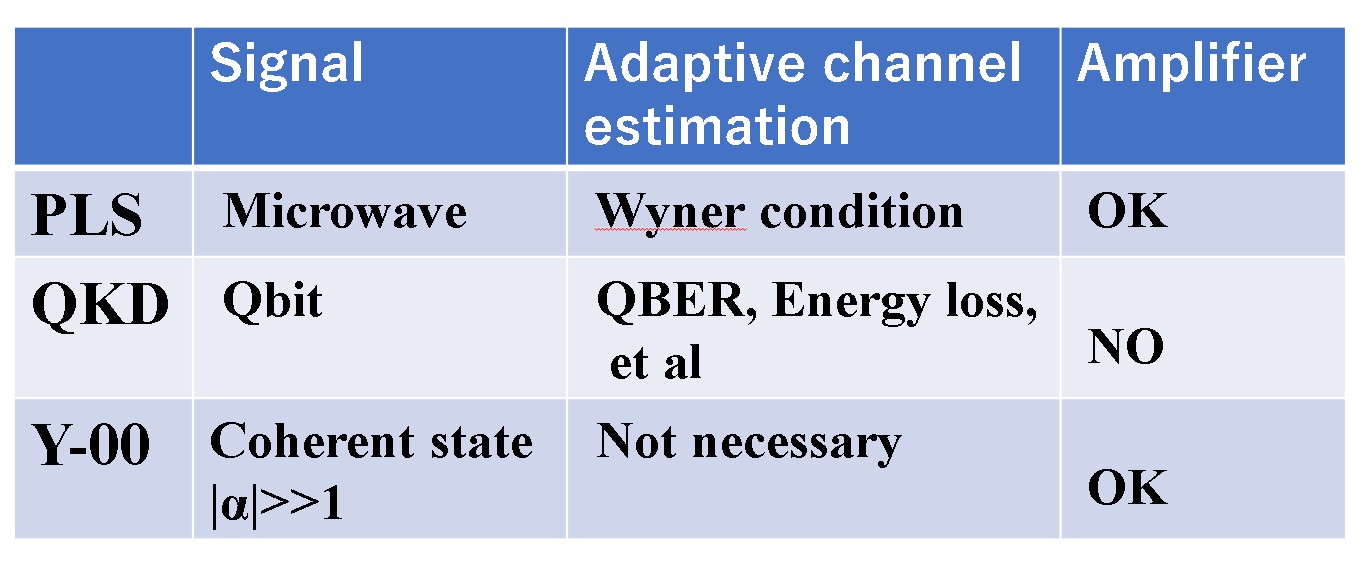}}
\caption{\textbf {Preliminary information and requirement for security metrics.} The PLS mechanism must constantly check 
whether the Wyner condition is met. In the case of QKD, the communication path becomes very complex, 
so a lot of prior information is required for security design. In the case of Y-00, it does not require.}
\end{figure}

\begin{figure}
\centering{\includegraphics[width=8cm]{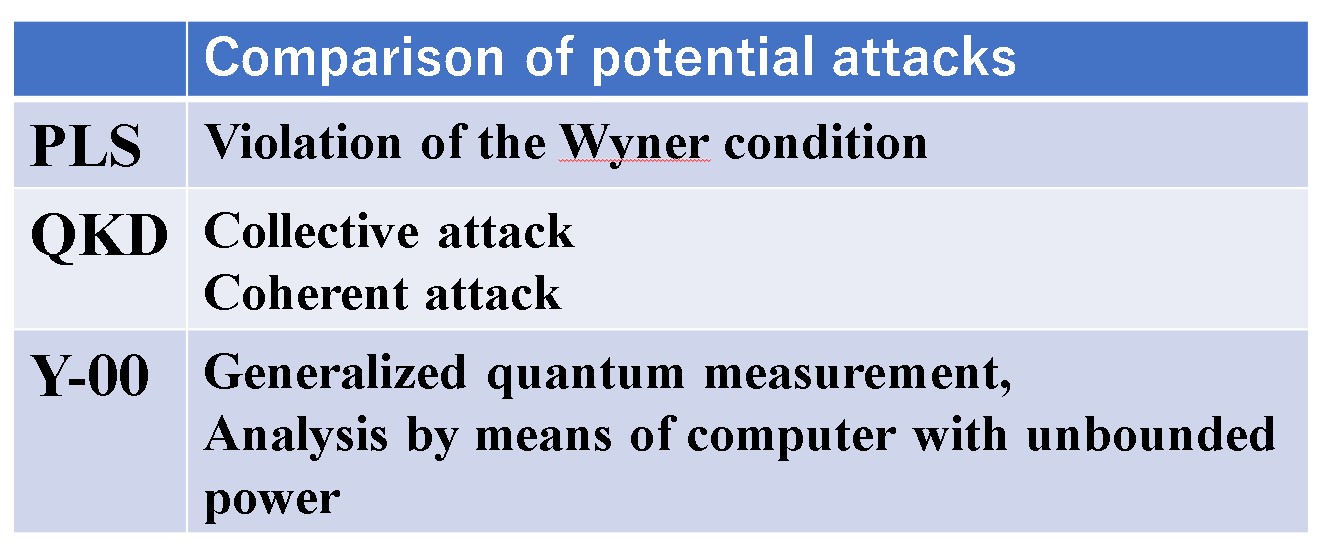}}
\caption{ \textbf {Comparison of attacks.} }
\end{figure}

\begin{figure}
\centering{\includegraphics[width=8cm]{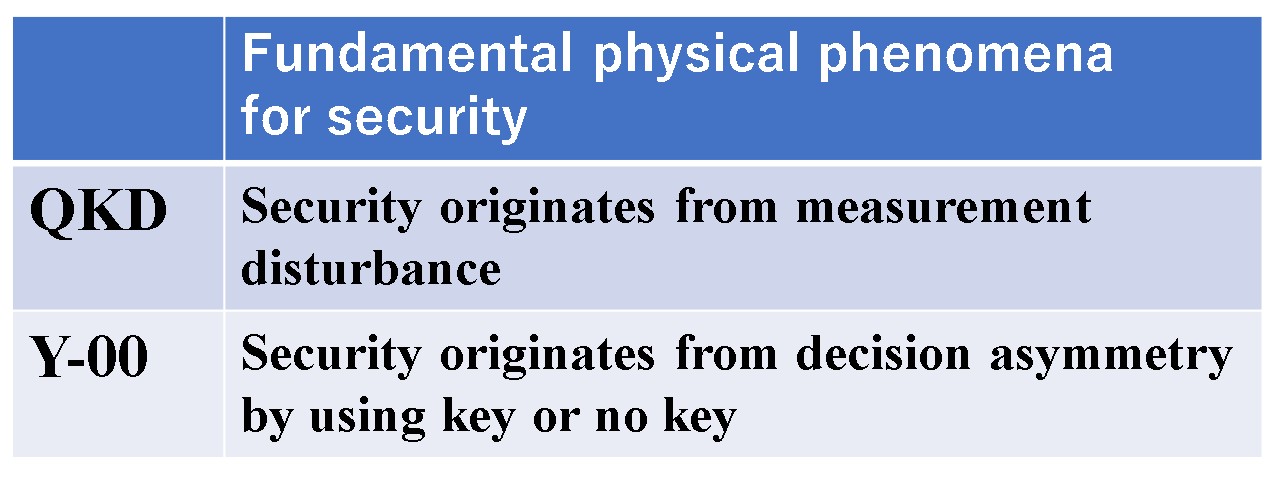}}
\caption{ \textbf {Physical phenomena for security.} In the case of QKD, the quantum phenomenon of origin of security is disturbance, 
and the communicator cannot control Eve's Holevo capacity itself.
In the case of Y-00, since Eve's quantum noise effect depends on the modulation mechanism, Eve's Holevo capacity can be controlled by design }
\end{figure}

\begin{figure}
\centering{\includegraphics[width=8cm]{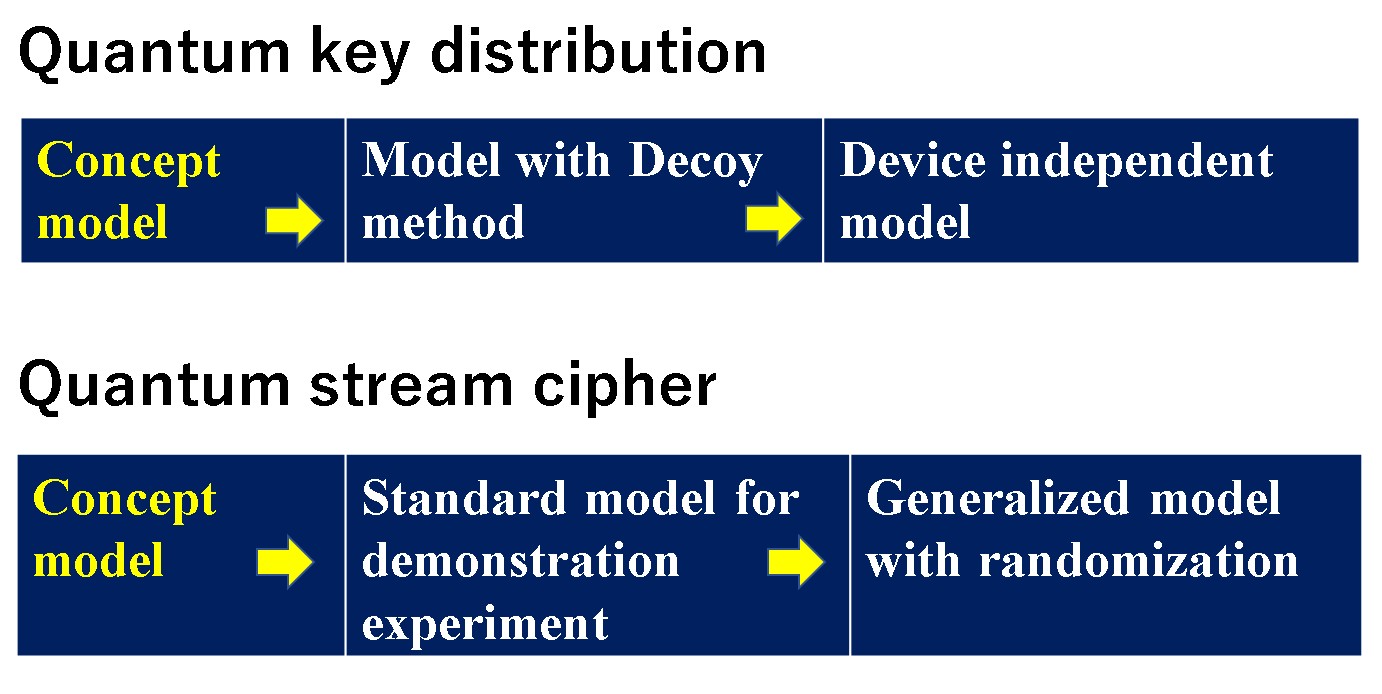}}
\caption{\textbf { Comparing the development history of QKD and Y-00.}}
\end{figure}

\end{document}